\documentclass[submission]{eptcs}
\providecommand{\event}{F-IDE 2021}
\usepackage{underscore}
\usepackage{listings}
\usepackage{amsfonts}
\usepackage{graphicx}
\usepackage{subcaption}

\title{Plotting in a Formally Verified Way}
\author{Guillaume Melquiond\thanks{This work was funded by the NuSCAP project (ANR-20-CE48-0014) of the French national research agency (ANR).}
\institute{Universit\'e Paris-Saclay, CNRS, ENS Paris-Saclay, Inria, Laboratoire M\'ethodes Formelles,\\ Gif-sur-Yvette, 91190, France}
\email{guillaume.melquiond@inria.fr}
}

\def\titlerunning{Plotting in a Formally Verified Way}
\def\authorrunning{G. Melquiond}

\begin{document}
\maketitle

\begin{abstract}
An invaluable feature of computer algebra systems is their ability to
plot the graph of functions. Unfortunately, when one is trying to design
a library of mathematical functions, this feature often falls short,
producing incorrect and potentially misleading plots, due to accuracy
issues inherent to this use case. This paper investigates what it means
for a plot to be correct and how to formally verify this property. The
Coq proof assistant is then turned into a tool for plotting function
graphs using reliable polynomial
approximations. This feature is provided as part of the CoqInterval
library.
\end{abstract}

\section{Introduction}

An invaluable feature of computer algebra systems (Maple, Mathematica,
etc) is their ability to plot the graph of mathematical functions.
Indeed, as the adage goes, a picture is worth a thousand words. When
encountering an unknown function, the first reflex of the user is to
plot it, so as to grasp its features. But how much can the user trust
that the plot is an accurate depiction of the graph of the function?

\begin{figure}[t]
\begin{subfigure}{0.495\textwidth}
\centering
\includegraphics[width=0.95\linewidth]{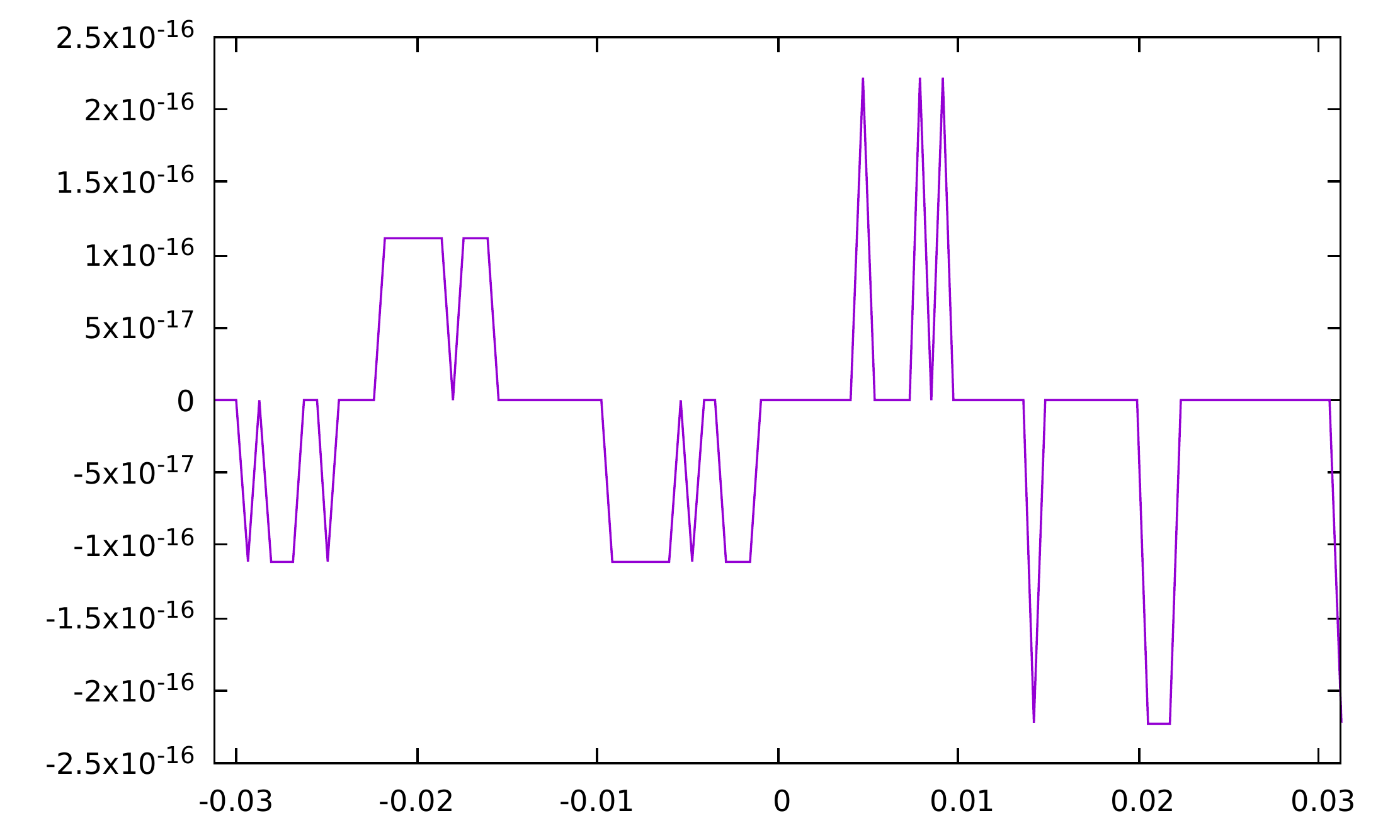}
\caption{Gnuplot}
\label{fig:bad}
\end{subfigure}
\begin{subfigure}{0.495\textwidth}
\centering
\includegraphics[width=0.95\linewidth]{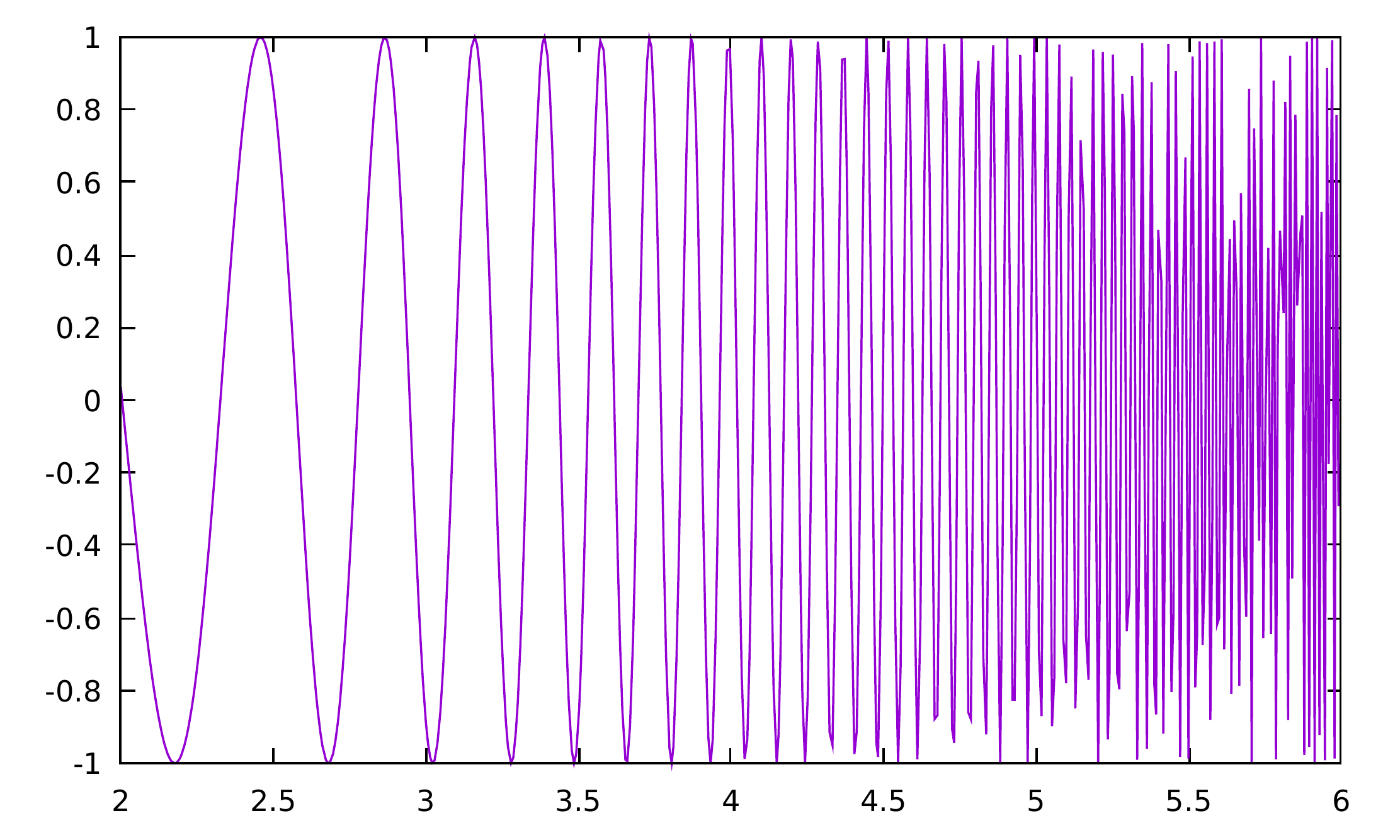}
\caption{Sollya}
\label{fig:rump-bad}
\end{subfigure}
\par\bigskip
\begin{subfigure}{0.495\textwidth}
\centering
\includegraphics[width=0.95\linewidth]{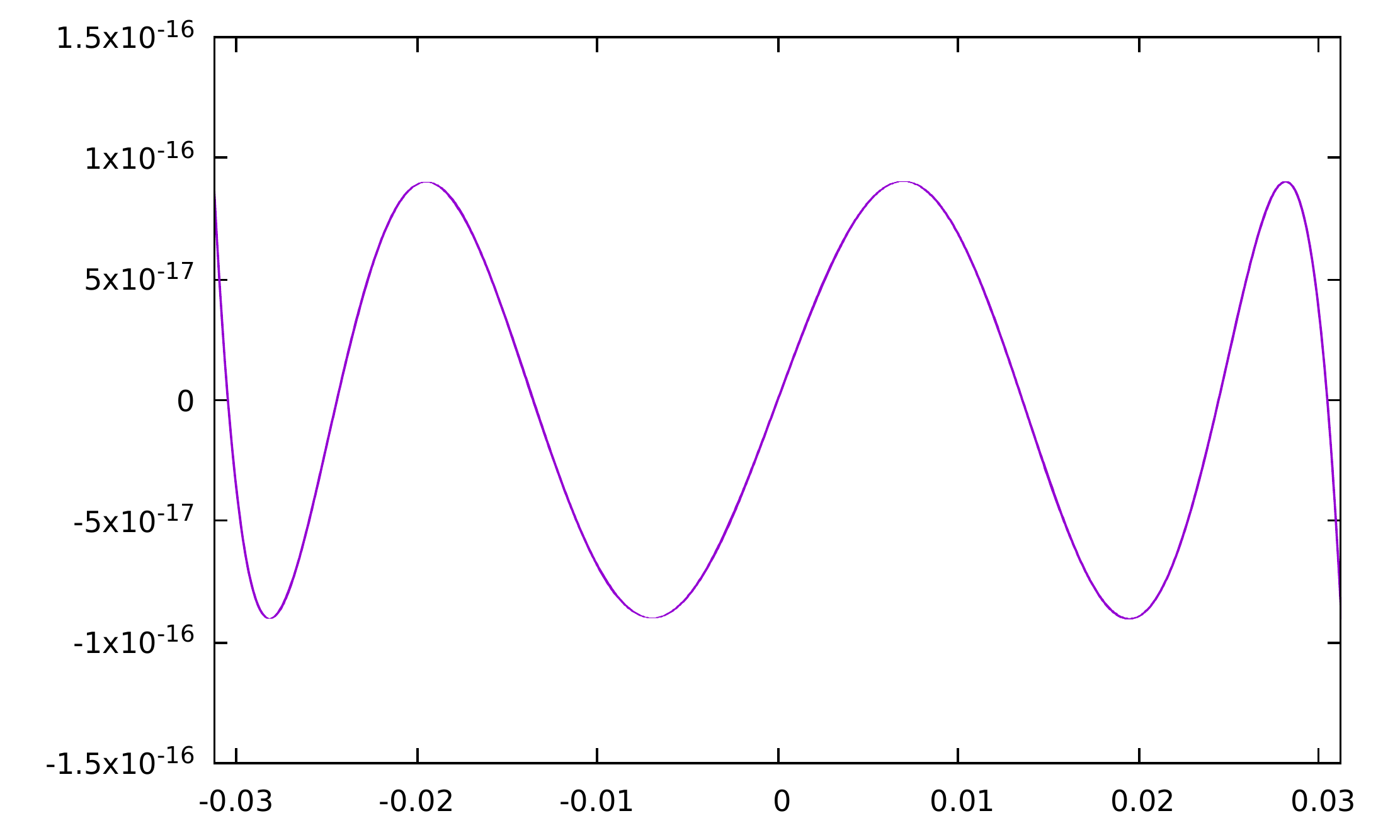}
\caption{Sollya and this work}
\label{fig:good}
\end{subfigure}
\begin{subfigure}{0.495\textwidth}
\centering
\includegraphics[width=0.95\linewidth]{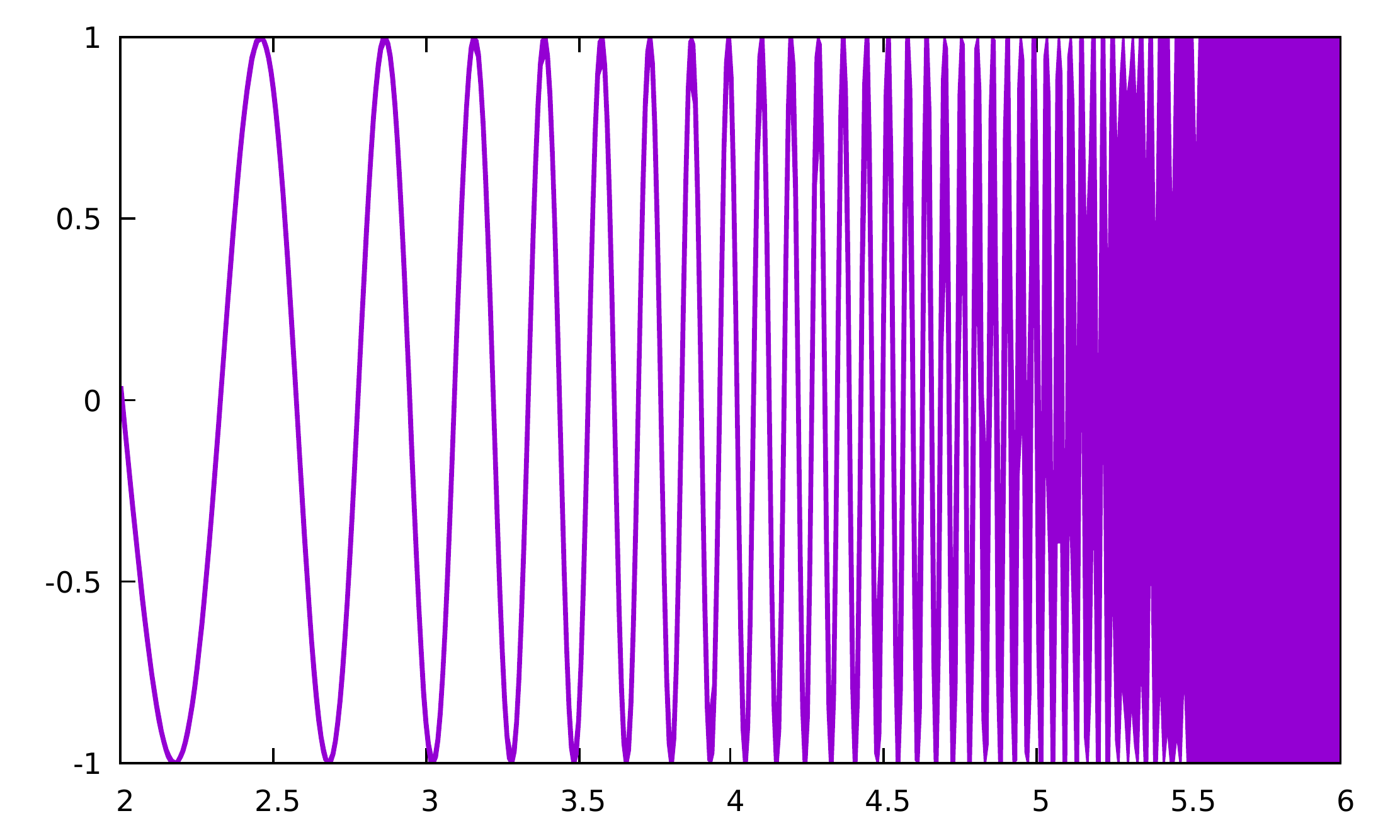}
\caption{This work}
\label{fig:rump-good}
\end{subfigure}
\caption{On the left, distance between $\exp$ and its minimax. On the right, $x \to \sin(x + \exp x)$.}
\end{figure}

Let us consider the case of a user who wants to implement
a floating-point library of mathematical functions. The implementation of
such a function, \emph{e.g.}, $\exp$, usually involves some polynomial $p$,
since processors are efficient at addition and multiplication. The
distance $|p(x) - \exp x|$ characterizes the quality of the approximation
and thus of the implementation. So, one might be tempted to plot $p(x) -
\exp x$ and look at its extrema. Let us assume that $p$ is
the minimax approximation of degree 6 between $-2^{-5}$ and $2^{-5}$ with
\emph{binary64} floating-point coefficients. Figure~\ref{fig:bad} shows
what the signed distance looks like, when plotted with Gnuplot. The
result certainly looks questionable, and other computer algebra systems
would hardly do better.

The main issue is that these systems plot the function graph using plain
\emph{binary64} arithmetic, which is way too inaccurate for our use
case. The Sollya tool was especially designed to solve this kind of
issue~\cite{CheJolLau10}. By using a 165-bit arithmetic, it is able to
produce Figure~\ref{fig:good}, which is representative of what the
distance between a function and its minimax polynomial usually looks like
(\emph{cf.} La Vall\'ee-Poussin's theorem). The user can now look at the
plotted function graph and see that the distance is bounded by
$10^{-16}$, which might be sufficient, depending on the purpose of the
mathematical library.

But what if the function graph needs more than 165 bits of precision to
be plotted correctly? Sollya actually performs its computations using
interval arithmetic. Instead of computing a single floating-point number
that approximates the real value of the function, it computes two
floating-point bounds that enclose this real value. So, by looking at the
distance between these bounds, it is able to detect when its internal
precision is not sufficient for a correct plot. In that case, the user
can increase the precision.

So, did Sollya actually solve the issue of plotting? Not quite. As other
computer algebra systems, it heuristically samples the plotted function.
So, any feature of the function that occurs between two
sampled abscissas will not appear on the plot. Consider the
function $f(x) = \sin(x + \exp x)$. This time, there is no precision
issue; even \emph{binary32} floating-point arithmetic could be used. But
the function oscillates so quickly that the plot is not faithful, as
shown on Figure~\ref{fig:rump-bad}. For $x \ge 4$, the function does not
even seem to reach $-1$ and $1$ anymore, although it is the sine function.

Let us define a \emph{correct plot} as follows: If a pixel is blank, then
there exists no value of $x$ such that $(x,f(x))$ falls into this pixel.
Conversely, a \emph{complete plot} is defined as follows: If a pixel is
filled, then there exists a value of $x$ such that $(x,f(x))$ falls into
it. This article explains how one can draw correct plots, as shown on
Figures~\ref{fig:good} and~\ref{fig:rump-good}. To increase the
confidence in the plots, they are performed using the Coq system. This is
not the first work to explore the topic of correct (and complete) plots
using a proof assistant~\cite{OCo08}, but this works focuses on producing
plots in a matter of seconds rather than
hours.\footnote{Figure~\ref{fig:rump-good} needs more than 600,000
  samples when using a method based on the modulus of uniform
  continuity~\cite{OCo08}.}

Section~\ref{sec:formal} explains how to formally define what a correct
plot is. Section~\ref{sec:plot} then shows how to compute it inside the
logic of Coq, using the CoqInterval library~\cite{MarMel16}. It also
explains how to get close to a complete plot. Coq gives the greatest
confidence in the correctness of the plots, but it hardly strikes as
a user-friendly system when it comes to computer algebra. So,
Section~\ref{sec:intf} focuses on interface concerns.

\section{Formal Specification}
\label{sec:formal}

The very first step is to state what it means for a plot to be correct.
We have some function $f$ from real numbers to real numbers. Note that
this is a function defined in a purely mathematical sense; no
floating-point numbers are involved here. We also have some bounds $x_1$
and $x_2$ between which we will plot the function graph. Again, these are
real numbers. We will not use them directly though, as it makes for much
more readable (and thus trustable) definitions to use two real numbers
$\mathit{ox}$ and $\mathit{dx}$ such that $x_1 = \mathit{ox}$ and $x_2 =
\mathit{ox} + \mathit{dx} \cdot w$. In other words, if the output device
has an horizontal resolution of $w$ pixels, then $\mathit{dx}$ will be
the width of a single pixel, while $\mathit{ox}$ will point at the left
border of the leftmost pixel.

A plot will then be defined as a list $\ell$ of intervals. The $i$-th
interval $\ell_i$ encloses all the possible values of the function for
the $i$-th pixel:
\[\forall x,~ \mathit{ox} + \mathit{dx} \cdot i \le x \le
\mathit{ox} + \mathit{dx} \cdot (i + 1) \Rightarrow f(x) \in \ell_i.\]
The translation to Coq is straightforward (with \texttt{I} a module
containing interval-related definitions):
\begin{lstlisting}
Definition plot1 (f:R->R) (ox dx:R) (l:list I.type) :=
  forall (i:nat) (x:R), ox + dx * i <= x <= ox + dx * (i + 1) ->
  I.contains (nth i l I.nai) (f x).
\end{lstlisting}

The interval \texttt{I.nai}, which contains any real number, is the
default value used when the list~$\ell$ is exhausted. Thus, the predicate
\texttt{plot1} is still valid when $i$ exceeds the length of~$\ell$. In
particular, if this length is smaller than $w$, the rightmost part of the
plot will be entirely comprised of filled pixels.

This could be the end of it, but since \texttt{I.type} is an internal
datatype, it might be difficult to turn the list into a bitmap or to
serialize it to another tool. So, in order to use a more universal
datatype, namely integers of type \texttt{Z}, a second predicate is
defined. To do so, we need two more real numbers: $\mathit{oy}$ and
$\mathit{dy}$. They play a role similar to $\mathit{ox}$ and
$\mathit{dx}$, but along the vertical axis:
\begin{lstlisting}
Definition plot2 (f:R->R) (ox dx oy dy:R) (h:Z) (l:list (Z*Z)) :=
  forall (i:nat) (x:R), ox + dx * i <= x <= ox + dx * (i + 1) ->
  oy <= f x <= oy + dy * h ->
  let r := nth i l (0, h) in
  oy + dy * (fst r) <= f x <= oy + dy * (snd r).
\end{lstlisting}

The seemingly counterproductive hypothesis $\mathit{oy} \le f(x) \le
\mathit{oy} + \mathit{dy} \cdot h$ is there in case the user wants to
focus on some detail of the function graph and is not interested in
extreme values that happen near it. As translating these extreme values
to integers is useless and possibly dangerous (due to overflow), only the
pixels with ordinates between $0$ and $h$ are kept, with $h$ the vertical
resolution of the output device.

If the user did not explicitly provide $y_1 = \mathit{oy}$ and $y_2 =
\mathit{oy} + \mathit{dy} \cdot h$, they can be found by computing the
union of all the intervals of a list satisfying \texttt{plot1}. In that
case, the hypothesis $\mathit{oy} \le f(x) \le \mathit{oy} + \mathit{dy}
\cdot h$ is trivially satisfied. This explains why the proposed mechanism
involves two predicates \texttt{plot1} and \texttt{plot2}, instead of
having just \texttt{plot2}. The following is an instance obtained for the
function $x \mapsto x^2$ between $0$ and $1$ for $w=10$ and $h=100$.
Notice how $\mathit{oy} \simeq -0.0003 \le 0$ and $\mathit{oy} + \mathit{dy} \cdot
h \simeq 1.015 \ge 1$.
\begin{lstlisting}
plot2 (fun x => x^2) 0 (820/8192) (-5/16384) (665/65536) 100
  ((0, 2) :: (0, 5) :: (3, 9) :: (8, 16) :: (15, 25) :: (24, 36)
   :: (35, 49) :: (48, 64) :: (62, 81) :: (79, 100) :: nil)
\end{lstlisting}

\section{Plotting a function graph}
\label{sec:plot}

Converting a list that satisfies \texttt{plot1} into a list that
satisfies \texttt{plot2} is a bit technical, but there is no difficulty
in formally verifying the algorithm using Coq. Thus, let us focus on getting
a list of intervals that satisfies \texttt{plot1}.

The first step is to turn the function $f$ into a straight-line program
that can be manipulated by functions written in Gallina, the language of
Coq, a dependently-typed lambda-calculus with inductive datatypes. To do
so, we just reuse the machinery of the \texttt{interval} tactic, that is,
an Ltac oracle reifies the function and Coq then formally checks that the
reified function matches the original one~\cite{MarMel16}. For example,
if the original function is $x \mapsto \cos x + 3$, the reified function
will look like \texttt{(Binary Add (Unary Cos (Var 0)) (Const 3))} where
\texttt{Binary}, \texttt{Cos}, etc are constructors of some inductive
datatypes.

The second step is to compute the list $\ell$ such that $f(X_i) \subseteq
\ell_i$ for $X_i = [\mathit{ox} + \mathit{dx} \cdot i; \mathit{ox} +
  \mathit{dx} \cdot (i + 1)]$. The natural idea would be to use interval
arithmetic to do so. Indeed, for every operation $\diamond$ over the real
numbers, it provides an operation over intervals (abusively noted
$\diamond$ too) that is compatible with it:
\[\forall U, V \in \mathbb{I},~ \forall u, v \in \mathbb{R},~
u \in U \land v \in V \Rightarrow u \diamond v \in U \diamond V.\]

\begin{figure}
\begin{subfigure}{0.495\textwidth}
\centering
\includegraphics[width=0.95\linewidth]{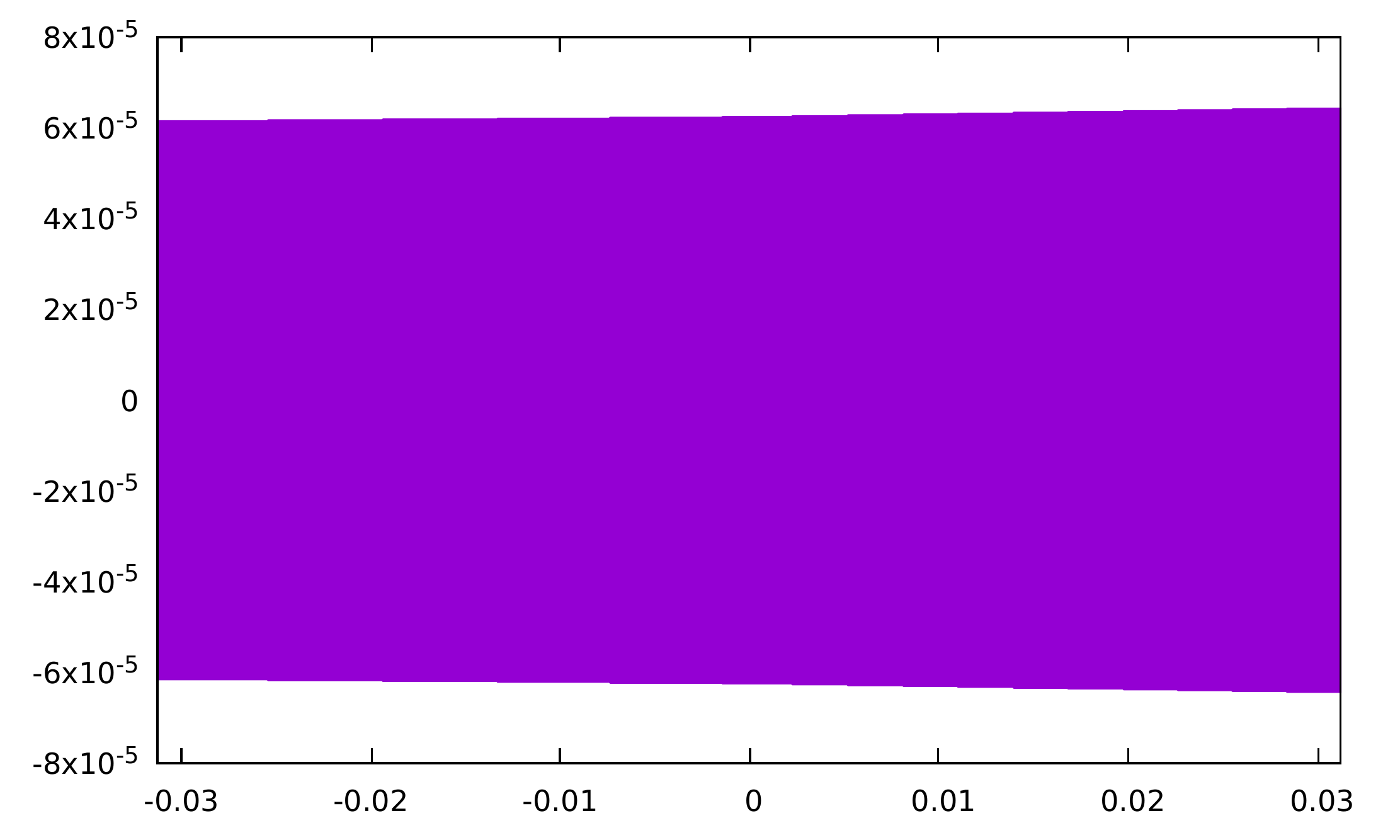}
\caption{Naive interval arithmetic.}
\label{fig:naive}
\end{subfigure}
\begin{subfigure}{0.495\textwidth}
\includegraphics[width=0.95\linewidth]{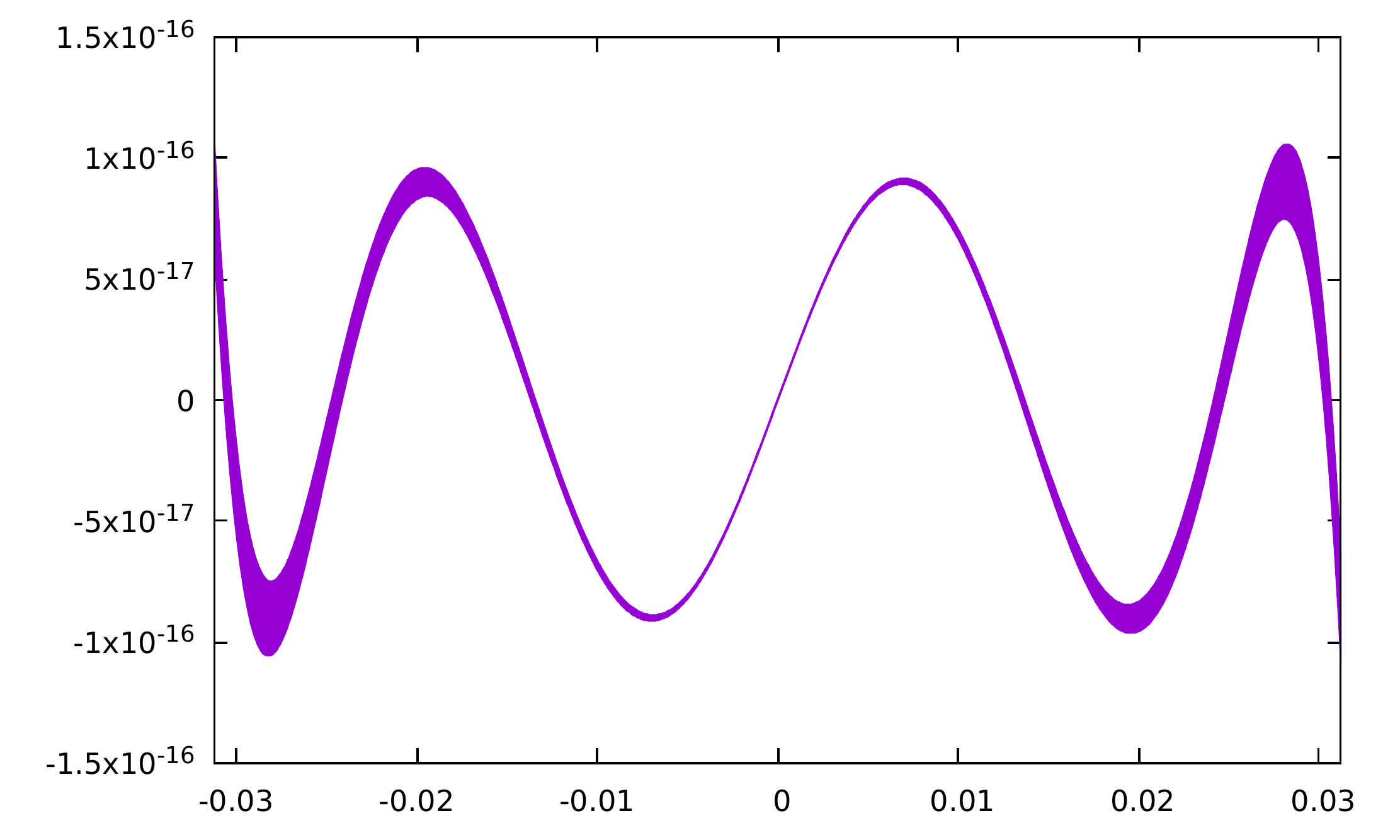}
\caption{Completeness check disabled.}
\label{fig:nocheck}
\end{subfigure}
\caption{Correct but hardly complete variants of Figure~\ref{fig:good}.}
\label{fig:incomplete}
\end{figure}

So, we would just have to recursively visit the reified expression,
applying the corresponding interval operations along the way. This would
give us some interval $\ell_i$ that satisfies \texttt{plot1}. While it
might work most of the times, it is not suitable for the use case
described in the introduction. Indeed, the core defect of interval
arithmetic, \emph{i.e.}, loss of correlation, applies here. A pixel might
be too wide for interval arithmetic to produce a meaningful interval. In
other words, $\ell_i$ will contain $f(x)$, but it will also contain many
other ordinates. Figure~\ref{fig:naive} shows the result; not only is it
a massive block of pixels, but the bounds are off by several orders of
magnitude: $6 \cdot 10^{-5}$ instead of $10^{-16}$.

There is a second, more practical, reason for not using naive interval
arithmetic. Since we are performing all our computations in the logic of
Coq, evaluating $w=512$ instances of the interval implementation of cosine
might take too long for an interactive use of Coq.
The solution is to compute a rigorous polynomial approximation
$(p,\Delta)$ of $f$ over some large interval~$X$, ideally
$W = [\mathit{ox};\mathit{ox} + \mathit{dx} \cdot w]$:
\[\forall x \in X,~ p(x) - f(x) \in \Delta.\]
The idea of using these polynomial approximations originated from
Sollya~\cite{CheJolLau10}. They were later formalized in
Coq~\cite{MMPRT13}. Eventually, they joined the CoqInterval
library~\cite{MarMel16}.

These polynomial approximations were instrumental when devising the
\texttt{integral} tactic for guaranteed numerical
quadrature~\cite{MahMelSib19}. Indeed, once $(p,\Delta)$ has been
computed, one can easily enclose the integrals of $p(x)$ and of $p(x) -
f(x)$ over~$X$, and thus the integral of $f(x) = p(x) - (p(x) - f(x))$
over~$X$. Since the ability to numerically integrate a function is not
that different from the ability to accurately plot its graph, we follow
a similar approach here. Once $(p,\Delta)$ has been computed, we use it
to compute an enclosure $Y_i$ of $f(X_i)$. This time, we can use
naive interval arithmetic to evaluate $Y_i = p(X_i) + \Delta$.

If $p$ has degree $0$, then the result is similar to the one obtained
using naive interval arithmetic. With degrees $1$ and $2$, the plot is
still an indiscriminate block of pixels, though the bounds are less
overapproximated. With degree~$3$, losses of correlation at the pixel level
are completely accounted for. So, the plot looks fine, but computing it
is way too slow for interactive use. Indeed, a single polynomial is not
sufficient for the whole interval $W$, since there is no way a degree-3
polynomial could ever meaningfully approximate a function with 7~roots.
So, the interval $W$ has to be split into many subintervals $X$, on each of
which $p$ has to be computed. The best running time is obtained for
degree 6. Then, the higher the degree, the slower it gets. Indeed,
decreasing the number of subintervals no longer compensates the
increasing cost of computing $p$ and evaluating it for every pixel. The
optimal degree highly depends on the plot, so its choice is left to the
user. By default, degree 10 is used, as with tactic \texttt{integral}.

Thanks to the rigorous polynomial approximations, we now have a correct
plot that is formally verified. But as shown on Figure~\ref{fig:naive},
a correct plot is not necessarily a complete one. So, to increase the
usability of our approach, we would like the plot to never be more than
a few pixels wide. A first idea would be to measure the width of
$p(X_i)$. If it is larger than a few pixels, then $X$ needs to be
subdivided further. Unfortunately, this causes too many subdivisions when
the function varies quickly, which is the case at the left and right ends
of Figure~\ref{fig:good}. A second idea would be to measure the width of
the error interval $\Delta$ to decide whether $X$ is sufficiently small.
Unfortunately, this still does not work. Indeed, the further from the
center of $X$, the worst the loss of correlation becomes when evaluating
$p(X_i)$, to the point where it becomes noticeable, as shown on
Figure~\ref{fig:nocheck}. So, this time, there are not enough
subdivisions.

To strike a balance between these two issues, the code computes an
underestimation of $f$ over $X_i$ and compares it to the overestimation
$Y_i = p(X_i) + \Delta$. If the latter is only a few pixels larger than
the former, then it is deemed good enough. Concretely, the code computes
$Z = p(\mathrm{lower}(X_i)) + \Delta$. There is a value of $x \in X_i$
such that $f(x)$ lies between the lower bound of $Y_i$ and the upper
bound of $Z$. So, if the distance between these two bounds is smaller
than a few pixels, the lower bound of $Y_i$ is accurate enough. If not,
the code tries again with the upper bound of $Z' = p(\mathrm{upper}(X_i))
+ \Delta$. If the lower bound of $Y_i$ is accurate enough, the code then
checks the upper bound by comparing it to the lower bounds of $Z$ and
$Z'$.

To finish, there is some kind of a chicken-or-egg problem. If the user
has not provided $\mathit{dy}$, how does the code know the height of
a pixel used to check for pseudo-completeness? To estimate it, the code
samples the function at 50 uniformly spaced points of $W$. This gives an
underestimation of $[y_1;y_2]$ and thus of $\mathit{dy}$. If the sampled
values do not capture the extreme values of the function, then the
predicted value of $\mathit{dy}$ is too small, which causes the plot to
be uselessly accurate and thus slower to compute.

\section{Interface}
\label{sec:intf}

We now have an algorithm (run inside the logic of Coq) that, given some
reification of function $f$ and some values for $\mathit{ox}$,
$\mathit{dx}$, etc, computes a list $\ell$ of pairs of integers. We also
have a theorem formally verified in Coq that states \texttt{(plot2 $f$
$\mathit{ox}$ \ldots~$\ell$)}. So, only interface issues remain.

First, let us deal with the plot display. The list $\ell$ is almost
a run-length encoding of the function graph, assuming a column-major
order. Indeed, a pair $(y_1,y_2) \subseteq [0;h]$ of integers represents
a column of first $y_1$ blank pixels, then $y_2 - y_1$ filled pixels, and
finally $h - y_2$ blank pixels. Getting Gnuplot to draw the resulting
bitmap is easy. Unfortunately, it is difficult to make sure that Gnuplot
maps one pixel of the bitmap to exactly one pixel of the screen. As
a consequence, some features of the plot might disappear if the drawing
area is just one pixel too small. Conversely, if the user tells Gnuplot
to zoom in (or just enlarges the drawing window), then the plot starts
looking blocky.

So, rather than a bitmap, it is visually more satisfying to turn the plot
into two piecewise affine curves that enclose the filled pixels of the
bitmap. This vector encoding allows the user to freely zoom on the plot
or resize the Gnuplot windows. Computing these two curves is actually
quite easy. Given two consecutive elements of the list $\ell_{i-1} =
(y_1,y_2)$ and $\ell_i = (y'_1, y'_2)$, one just needs to associate
to abscissa $\mathit{ox} + i \cdot \mathit{dx}$ the ordinates
$\mathit{oy} + \min(y_1,y'_1) \cdot \mathit{dy}$ and $\mathit{oy} +
\max(y_2,y'_2) \cdot \mathit{dy}$. The band between the two curves
contains all the filled pixels of the original bitmap, thus guaranteeing
the correctness of the plot, at the expense of being a bit less narrow
than the bitmap one.

As for the user queries, let us take some inspiration from existing
computer algebras system. They often handle plots as first-class citizen,
that is, the user can execute ``\texttt{p := plot(f,x1,x2)}'' to store
a function graph into some variable \texttt{p}. Then, simply executing
\texttt{p} causes the plot to be displayed. (Both steps can usually be
merged into a single one, if the function graph does not need to be
stored for later use.) We can follow a similar approach for Coq.
Unfortunately, Coq requires top-level terms to be preceded by a command,
\emph{e.g.}, \texttt{Print}, \texttt{Check}, \texttt{About}. We cannot
reuse an existing command, so we add yet another one: \texttt{Plot p}.
This causes Coq to open a Gnuplot windows using the data encoded in the
\texttt{plot2} type of \texttt{p}. As for a Coq equivalent to
``\texttt{p := plot(f,x1,x2)}'', we combine the \texttt{Definition}
command with the tactic-in-term feature of Coq. This provides the
following interface: \texttt{Definition p := ltac:(plot f x1 x2)}. The
\texttt{plot} tactic can take two extra arguments to specify the ordinate
range. When absent, the tactic computes the extrema of the function
between the endpoints.

The \texttt{plot} tactic supports the same configuration mechanism
as the tactics \texttt{interval} and \texttt{inte\-gral}~\cite{MarMel16}.
For instance, to obtain Figure~\ref{fig:good}, one needs to increase the
precision to 90 bits, as follows:
\begin{lstlisting}
plot (fun x => 1+x*... - exp x) (-1/32) (1/32) with (i_prec 90)
\end{lstlisting}

A new flag has been added to specify the dimension of the bitmap:
\texttt{i\_size w h}. By default, the tactic produces a plot of size
$512\times384$. Other meaningful flags are \texttt{i\_degree} to control
the degree of the polynomial approximations and
\texttt{i\_native\_compute} to tell Coq to first compile the algorithm to
machine code rather than directly interpreting it. This might be useful
to speed up some computationally-intensive plots. Indeed, the
architecture of Coq unfortunately forces the tactic to execute the algorithm twice:
once to get the actual list and a second time to instantiate the
correctness theorem.

\section{Conclusion}

This article has presented a mechanism integrated in release 4.2 of the
CoqInterval\footnote{\url{https://coqinterval.gitlabpages.inria.fr/}}
library. It makes it possible to compute formally correct function graphs
and display them directly from Coq. Despite the computations being
performed inside the logic of Coq, performances are good enough for
interactive use. For example, it takes less than 4 seconds to compute and
formally verify the complicated plot of Figure~\ref{fig:rump-good} with
the default settings, and about 1 second with
\texttt{i\_native\_compute}.

While the current interface is a bit unfriendly, we could readily imagine
a new front-end that would exempt the user from typing
\texttt{Definition} and \texttt{Plot}, as well as the \texttt{ltac:(...)}
quotation mechanism. The long-term goal is to revisit the
way Coq is used, making it more of a computer algebra system,
\emph{e.g.}, through interfaces such as CoCalc and
Jupyter~\cite{PerGra07}. Plots would no longer be opened in separate
windows but directly embedded in the document. In the meantime, by virtue
of the plotting algorithm being written in Gallina, it could easily be
extracted to OCaml and distributed as a standalone library.

\newpage

\bibliographystyle{eptcs}
\bibliography{biblio}
\end{document}